# Multiparty Authorization for Secure Data Storage in Cloud Environments using Improved Attribute-Based Encryption

Partha Paul[1,] Keshav Sinha[2]

[1]Birla Institute of Technology, Mesra, India, [2]School of Computer Science, UPES Dehradun, India

**ABSTRACT**

In today's scenario, various organizations store their sensitive data in the cloud environment. Multiple problems are present while retrieving and storing vast amounts of data, such as the frequency of data requests (increasing the computational overhead of the server) and data leakage while storing. To cope with said problem, Attribute-Based Encryption (ABE) is one of the potential security and access control techniques for secure data storage and authorization. The proposed work divides into two objectives: (i) provide access to authorized users and (ii) secure data storage in a cloud environment. The improved ABE using Functional Based Stream Cipher (FBSE) is proposed for data storage. The proposed technique uses simple scalar points over a parabolic curve to provide multiparty authorization. The authorization points are generated and share only with the authorized recipients. The Shamir secret sharing technique generate the authorization points and 2D-Lagrange Interpolation is used to reconstruct the secret points from regular parabola. The proposed scheme has specified the threshold ($T_s \geq 3$) legally authorized users to reconstruct the attribute-associated keys for decryption. The encryption of data is evaluated using Statistical analysis (NIST Statistical Test Suite, Correlation Coefficient, and Histogram) test to investigate image pixel deviation. The parameters like encryption and decryption are used for performance analysis, where an increase in the number of attributes for the authorization policy will increase the encryption time. The proposed scheme imposes minimal storage overhead, irrespective of the user's identity. The security analysis evidence that it resists collision attacks. The security and performance analysis results demonstrate that the proposed scheme is more robust and secure.

## 1. INTRODUCTION

In the modern era, the cloud has become the destination platform for data storage. At the same time, handling data storage and management has become challenging for any large organization. Data owners tend to outsource their massive chunks of data on cloud servers with some pre-designed Access Control Policy (ACP) [2]. The cloud seamlessly delivers various resources and services, readily available whenever needed, while imposing a marginal cost, ensuring optimal flexibility and economic efficiency [1]. In a cloud environment, the data owners have limited authority over stored data. The traditional cryptographic technique is used to ensure stored data security. The traditional cryptosystem uses complex key management (such as Random Number Generation (RNG), Key Derivation Functions (KDF), Key Exchange Protocols, Key Wrapping, and Key Pair Generation) with high storage overhead [3]. In response to encryption challenges, researchers have turned to utilizing Attribute-Based Encryption (ABE) as a viable solution. ABE empowers them to tackle encryption issues by enabling access control based on predefined attributes or characteristics, thus enhancing the flexibility and granularity of data protection. The

ABE is the public-key cryptosystem technique that uses access control and authorization policy for data sharing. The study by Sahai and Waters (2005) [24] explored the intricate interplay between users and stored data, shedding light on their relationship dynamics. The encryption scheme was formulated as a representation of the identity (I) and its ability or inability (θ) to operate on a Resource (R) within the ecosystem, encompassing all possible combinations of 'I', 'θ', and 'R' [4]. ABE has several disadvantages, such as substantial computation costs, encryption/decryption time, and key management [5]. The fined-grained Access Control List (ACL) is used to overcome the mentioned problem. In today's scenario, few systems have been implemented based on ACL and ABE in cloud environments [7]. However, using ABE provides flexibility but increases overhead during multi-linear mapping and public key transmission [8]. This study aims to minimize encrypted data access and solely provide the decryption key to authorized individuals. The system is designed based on three factors (i) Key Management, (ii) Authorization Policy, and (3) Privacy Prevention. The key management works on the Access Control List (ACL), which uses the functional-based Stream cipher for data encryption/decryption, and improved RSA uses for secure key sharing. The authorization policy uses simple scalar points over a parabolic curve for key generation. The access control policy will update at every ΔT sec to verify additional attributes, which helps to protect confidential data leakage. The key contribution of the proposed system is summarized as follows:

- The access control scheme and ABE are used for secure data storage and sharing.
- FBSE encrypts data and stores it on the organization's server.
- The parabolic equation uses owner attributes to create a key distribution system.
- Lagrange interpolation is used to reconstruct the decryption key.
- Users only access and decrypt data if they have access privileges.

The paper is organized as follows: Section 2 depicts the access control and revocation of privileged user access; Section 3 presents the framework for key sharing and encryption. Section 4 presents the image encryption performance and security analysis. Finally, the entire work is concluded in Section 5.

## 2. ACCESS CONTROL AND REVOCATION

In a cloud environment, the role of the cloud service provider (CSP) is to manage the storage and sharing of data. Data security is determined based on cryptographic techniques and Access control mechanisms [24]. An encryption technique is used to encrypt data and securely store it on a server. The encryption key is only shared with authorized users [9]. On the other hand, access control schemes ensure that only authorized individuals will gain entry or interact with specific resources and restrict access to unauthorized users.

### 2.1 Access Control on Encrypted Data

An access control scheme is implemented on encrypted data to restrict unauthorized access [6]. The problem with this approach is trust and the relationship between data owner and users. Trust and relationships change dynamically in different scenarios, and to maintain the desired level of security, the decryption key must be updated regularly [11]. The access control list grants user access instead of sharing the symmetric key with all participants. However, this method has a drawback, such as (i) it creates various groups of files and (ii) the size of the data owner's key increases [12]. The alternative approach is to combine symmetric and public-key cryptography. The data owner shares the public key with authorized users, and decryption is performed using the access control list [13]. However, as the number of users increases, it becomes expensive and time-consuming. The effective way is to encrypt data and provide the necessary keys to users at every ΔT time interval, allowing them to decrypt the data only with the authorization of the data owner [10]. Attribute-Based Encryption (ABE) has recently undergone significant developments, where it recognizes users based on attributes rather than a single identity. In the ABE system, data is encrypted using a specified access control list. The authentication is performed based on user attributes matching the ACL to decrypt the data [14]. Recent literature has demonstrated that Attribute-Based Encryption (ABE) is extensively utilized for secure data storage in cloud computing [15]. However, it is typically complex and requires substantial computation, making it challenging to implement in practical situations. Lai et al. [16] introduced Shamir's Secret Sharing algorithm to maintain confidentiality and integrity in a multi-cloud environment. It involves verification of each participant's secret share point, which is used for authorization. The hash-based signature shares generate the point for the authorized personnel. Yang et al. (2013) [17] introduced a data access control scheme named DAC-MACS. It employs an efficient access control list to provide authorization for users and security for storing data in the cloud environment. The approach uses the decryption and revocation of the user's permission by using trust between the user and the data owner. DAC-MACS uses the independent global certificate authority (CA) and multiple attribute authorities (AAs) to authorize and identify global users and overcome the issue of collusion attacks.

### 2.2 Permission Revocation for Data Owner

The data owner withdraws access permissions from individuals in the access control list who no longer maintain group affiliation or due to a lack of trustworthiness on the user's part [18]. User revocation is a well-studied yet challenging problem. The main concern is that revoked users still have access to the previously mentioned keys, allowing them to decipher ciphertext. As a result, anytime a user is revoked, the data owner must perform re-keying and re-encryption activities to prevent the revoked user from accessing future data [19]. ABE uses data owner attribute to encrypt the data and distribute keys to authorized users regularly. Due to the substantial burden conducted by the data owner, this strategy is inefficient [22]. Allowing the data owner to delegate computationally costly processes, such as re-encryption, to a third party while exposing the least amount of information is a superior approach. Proxy Re-Encryption (PRE) is a valuable option since it allows a semi-trusted proxy to transform a ciphertext that decodes into another ciphertext. It is decrypted without knowing the underlying data or using secret keys [21]. To integrate the KP-ABE with PRE to outsource most of the calculation operations involved in user revocation to the CSP. It first integrates PRE with a CP-ABE system in the cloud to provide a scalable revocation mechanism. The attribute revocation is supported by the work stipulating that if a user is removed from a system, the data owner must transfer PRE keys to the CSP [23]. The biggest flaw with this solution is that it requires the data owner to be online to submit the PRE keys to the CSP on time, preventing the revoked user from accessing the data. The delay in issuing PRE keys might endanger system security [20]. The proposed model solves the problem by introducing a reputation center representing the data owner and in charge of re-encryption. The reputation-based insurance mechanism decreases the risks of data exposure after access. Meanwhile, owing to large-scale data handling, CSP's computational burden should be minimal to ensure optimal efficiency. The proposed model uses the re-encryption process of encrypted data, whereas the re-encryption key is not stored in the cloud servers to reduce the CSP's re-encryption process load [31].

# 3. ENCRYPTION SCHEME AND KEY SHARING

This section presents the Access Control List (ACL) based Attribute-based Encryption technique for encrypting and sharing data. There are several objectives for the implementation of the proposed framework, such:

- The access control list may struggle with scalability, and to overcome the problem ABE approach is used to meet security and privacy requirements.
- The proposed framework allows users to access the encrypted data only if the encryption attributes match the k-attributes of authorized users.
- The link between the attribute sets, user credentials, and open attributes potentially compromises data privacy.
- We are modifying the access control and performing the re-encryption process on the ciphertext.
- Reputation Center (organizational server) is responsible for the re-encryption of data.
- The CSP is not responsible for storing the encryption key.

The reputation-based insurance mechanism mitigates data exposure risks and minimizes the CSP computational liability.

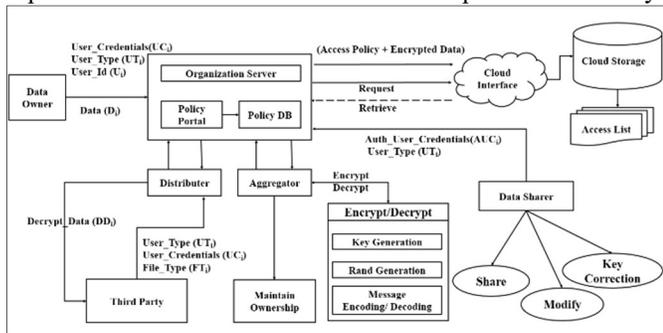

**Figure 1. Distributed Access Control Scheme Model**

Fig. 1 presents the framework for a distributed organizational environment where data is shared across several users. The proposed system consists of four participants (i) Data Owner, (ii) Cloud Storage, (iii) Data User, and (iv) Organization Server.

- A data owner is a registered organization user who creates a file and Access Control List (ACL) and shares the file with other users. The owner is authorized to perform read/write and update/delete operations on the file and access control list (ACL). The attributes are User_Id (Ui), User_Type ($UT_i$), and User_Credentials ($UC_i$). Data sharing requires authorization credentials which are issued by the organization server.
- Cloud Storage is used to store encrypted data and backup ACL. It is considered a semi-trusted platform to perform read/write operations on ACL at every encrypted data request. The organization server requires an ACL to give authorization for every user. The cloud uses the repository structure to organize the data files.
- Data Users are registered users whose credentials are stored in the organization's server access control list. The organization's server is responsible for generating the multiparty authorization point using the parabolic curve and sharing it with the respective users for accessing the encrypted data. The attributes are User_Id ($U_i$), User_Type ($UT_i$), and User_Credentials ($UC_i$). It also keeps the sensitive data derived from authorized credentials.
- Organization Server acts as an authorization center between the data owner and users. It is responsible for file encryption and decryption, generating and sharing authorization points. The access control list of each file is stored and managed by the organization server. It also assigns, revokes, and manages the User_Credentials ($UC_i$) from the ACL. It also records and indexes all the valid credentials to avoid redundancy.

In the proposed framework there are several steps are required to perform functional operations (data owner, organization server, and cloud storage) as follows:

**Algorithm 1:** Step-by-Step process of the data owner, organization server, and cloud storage

**Step 1.** The data owner gets registered with the organization server, and the sensitive credentials are stored in the ACL.

**Step 2.** Data ($D_i$) share with the organization server.

**Step 3.** Policy_Portal in the organization server is used to create the policy based on the user's attributes and sharers. The Policy_DB is used to store each file policy based on priority.

**Step 4.** The work of the aggregator is to perform encryption and decryption based on the FBSC. Here, random padding is performed on the data to create redundancy.

**Step 5.** Encryption is performed based on the padded data's symmetric key cryptosystem (Functional-Based stream cipher). The symmetric secret key is re-encrypted using an asymmetric technique.

**Step 6.** The re-encrypted key is given to the parabolic equation to generate multiparty authorization points and share them with the authorized users.

**Step 7.** The decryption of data is performed based on the Shamir Secret Sharing technique. The secret key generates the user's attributes and authorization points.

**Step 8.** The data sharer uses the Auth_User_Credentials ($AUC_i$) and User_Type ($UT_i$) to share, modify, and make key corrections. The authorized users are only responsible for key corrections and modifications. The data owner has access to share the data and authorization keys.

## 3.1 System Functionality

This section presents the description of the proposed distributed access control scheme model. The primary functions are (i) Random Padding, (ii) Key Management, and (iii) Access Control List Policy.

### 3.1.1 Random Padding

The random bit generated using the KM Generator [26] is expressed as Eq. (1).

$$Seed_{n+1} = (Seed_n \times M \times I) \bmod n \quad (1)$$

I = Non-Integral Number, M = Maddy Constant, n = Moduli. The sequence is generated in the range of {0 to n}. The generated random sequence is probabilistic and stored in the variable ($N_k$). The mathematical proof is investigated in the paper of Sinha et al. (2022), and an extended application of the generator is used here for optimal asymmetric encryption padding for data. The fixed random bit [$N_k \rightarrow N_1$] and repeated random bit [$N_k \rightarrow N_2$] are generated based on the $N_k$. The generated random bits [$N_1, N_2$] are padded with the original message at ΔT time intervals.

### 3.1.2 Authorization Key Management

The proposed model uses two distinct types of keys (i) Data Owner Secret Key and (ii) Multiparty Authorization Key. The organization server uses the hierarchical key structure for data encryption and sharing of keys with different participants. The aggregator module is used to manage the symmetric and asymmetric keys. Data encryption uses the Functional-Based Stream Cipher, whereas the Improved RSA technique encrypts the symmetric key. The top-level key associated with the data owner attribute is usually stored with the ciphertext. The work of the organization server is to hold the symmetric key in the hierarchical structure. The top-level keys are managed in the system directory. The file system's directory structure is used to manage the multiparty authorization keys, whereas one key is used for one file.

### 3.1.3 Access Control List Policy

The organization server consists of a Policy_Portal and Policy_DB module to maintain the ACL for each user's file. The participant's Secret Key and authorization points are required to access the appropriate data from cloud storage.

## 3.2 Functional-Based Stream Cipher

The data is encrypted using a Functional-Based Stream Cipher (FBSC) before storing it on the cloud server. The traditional cryptosystem requires huge computational time, a complex permutation-combination system, and is vulnerable to cryptanalytic attacks. In the proposed model, the FBSC uses the Involutory function for key generation and overcomes the problem of the traditional cryptosystem. The involutory function uses bijective mapping (one-to-one correspondence). The involutory function's mathematical background is based on a bijective function where Set A and Set B consist of the same number of elements and the inverse function from Set B to Set A. The function satisfies the property f(AB) = f(B) f(A) for all {A, B} in the group G, where G is a group, making it an Anti-homomorphic function. The involution function is expressed as Eq. (2).

$$f(x) = X \quad (2)$$

And the inverse of Eq. (2) is expressed as Eq. (3).

$$f(f(x)) = X \quad (3)$$

The inverse involution function is also known as Anti-involution (or Anti-homomorphism). A similar mechanism applies to encryption, where a symmetric key is used for encryption, and decryption processes are defined as Eqs. (4) and (5).

$$Encryption = D_{encrypted_i} = E_k(D_i) \quad (4)$$

$$Decryption = D_i = D_k(D_{encrypted_i}) \quad (5)$$

Here $D_i$ = data, $E_k$ = Encryption key, $D_k$ = Decryption key. The proposed encryption process uses the stream cipher for data input. The involution function takes the data as a stream for encryption. The proposed Involution Function-Based Stream Cipher for encryption is defined as Eq. (6).

$$Encryption = f(x) = \left(PK_{SK} - x^{1/R_n}\right)^{R_n} \quad (6)$$

We compute the anti-evolution function $f(f(x))$, which is generated by Eq. (7).

$$Decryption = f(f(x)) =$$
$$\left(PK_{SK} - \left((PK_{SK} - x^{1/R_n})^{R_n}\right)^{1/R_n}\right)^{R_n} = x \quad (7)$$

Here, '$PK_{SK}$' and '$R_n$' are the variables, '$x$' is the plaintext, $f(x)$ is the ciphertext, and [$PK_{SK}, x \in I^+$]. To fit with the formula with the cryptographic aspect, ($PK_{SK}$) represent a symmetric secret key, and $R_n$ = randomized power variable. The key storage is based on the data file, where one key is linked with one file. If the user changes the file name, the file's authorization ($FK_i$) is also varied in the mathematical representation as Eq. (8).

$$FK_i = \begin{cases} a_i \oplus x_i \\ a_i' \oplus x_i \end{cases} \quad (8)$$

where $a_i$ and $a_i'$ are the old and new secret keys, '$x_i$' is the full name for the file '$i$'. It encrypts the file with the secret key '$a$', with the owner attributes to authorize only the privileged users.

## 3.3 Multiparty Authorization Key Computation

This section presents the implementation of ABE-based multiparty authorization key generation for data sharing. The authorization key distribution is performed when the data owner wants to share the ciphertext with access privileges. It grants access to authorized users to decipher the secure data. The key sharing is based on Shamir's Secret Sharing Key distribution technique. The distribution of each key uses the 2-degree langrage interpolation function over the finite field '$Z_p$'.

### 3.3.1 Shamir Secret Sharing

It is ideal for the ($k, n$) threshold scheme, dividing the secret key '$PK_{SK}$' into '$n$' pieces of data [$PK_{SK_1}, PK_{SK_2}, \ldots, PK_{SK_n}$]. If the adversary knows any '$k$' or more '$PK_{SK_i}$' pieces, make '$PK_{SK}$' easily assessable. The reconstruction of secret '$PK_{SK}$' combines '$k$' pieces of the data and formulation for constructing a polynomial using Eq. (9).

$$f(x) = a_0 + a_1 x + a_2 x^2 + \cdots + a_{k-1} x^{k+1} \quad (9)$$

Here $a_k$ = secret key and user attribute token ∈ {$a_1, a_2, \ldots a_{k-1}$}. The key distribution is present in points where '$n$' users are present with the privilege access policies on the server. The server creates the user attribute tokes sets [$i$ = (1, 2, 3, …, $n$)] to retrieve [$i, f(i)$]. Every authentic user has non-zero integer point input to the polynomial. For every given subset of '$k$', '$a_0$' is obtained using 2-degree Lagrange interpolation. The secret key construction function is defined over the finite field $Z_p$ and is expressed as Eq. (10).

$$G(X) = (a_0 + a_1 X + a_2 X^2) \bmod p \quad (10)$$

The distribution of keys is based on the number of users present on the server with privileged access.

**Algorithm 2:** Key distribution using parabolic curve

01: Start
02: $D_X$ = number of privileged users
03: Point generation from polynomials: $D_{X-1}$ = (X, G(X))
04: Point calculation: Y = (G(X))
05: $D_{X-1}$ = (X, Y), where [X = 1, 2, …, $D_X$]
06: End

Each participant performs single point sharing (X, G(X)). Here the points start from the (1, G(1)) and not from (0, G(0)) because G(0) consists of secret key information [32]. The distribution of '$X_i$' key point is only to the privileged users and simplified by Distributor and Aggregator. The points are present in the hash formed (0, $KC_i$). Here, $KC_i$ is the file binding code of the '$X_i$', is calculated using Eq. (11).

$$KC_i = (Sec_{K_i}, f(x)) \bmod p \qquad (11)$$

The interpolation function of file '$X_i$' must pass at least three points to generate the secret key $PK_{SK}$. The Lagrange basis polynomials generate the points using Eq. (12).

$$l_j(x) = \prod_{m \neq j}^{0 < m \leq k} \frac{x - x_m}{x_j - x_m} \qquad (12)$$

The system assumption is that no two points '$x_j$' are the same, $(m \neq j)$ and $(x_j - x_m) \neq 0$. Let us consider an n-users present on the server with having the privilege access policy. The users have the instances set (i = 1, 2, 3, …, n). The distribution of the key in-between different users is given in Eq. (12). Every user has granted the points (a non-zero integer input polynomial). Now, for every given subset of decryption key 'k', we can obtain '$a_0$' using the interpolation.

*3.3.2 Key Generation with Parabolic Equation*

The organization server generates multiparty authorization points by applying the Shamir Secret Sharing technique over the attributes of the file owner. We use the parabolic equation to construct the key distribution expressed as Eq. (13).

$$F(x) = a_0 + a_1 X + a_2 X^2 \qquad (13)$$

Here, '$a_1$' and '$a_2$' are the attributes of the owner, and '$a_0$' is the secret key. Now,

$$F(X) = 94X^2 + 166X + 1234 \qquad (14)$$

Then Number of Privileged Users $D_x$= 6, and $X_{x-1}$= (X, F(X)) from the polynomial.

$$\begin{cases} Y = F(X) = 1234 + 166X + 94X2 \\ D_{x-1} = (X, Y) \end{cases} \qquad (15)$$

Here, [X = 1, 2, 3, 4, 5, and 6]. Now, solve Eq. (14) by using Eq. (15), and the value of '$D_0$' valve is presented in Table 1.

**Table 1: Each participant's single points ($P_x$)**

| $D_{x-1}$ | F ($X_n, Y_n$) | Y |
|---|---|---|
| $D_0$ | F ($X_0, Y_0$) | 1, 1494 |
| $D_1$ | F ($X_1, Y_1$) | 2, 1942 |
| $D_2$ | F ($X_2, Y_2$) | 3, 2578 |
| $D_3$ | F ($X_3, Y_3$) | 4, 3402 |
| $D_4$ | F ($X_4, Y_4$) | 5, 4414 |
| $D_5$ | F ($X_5, Y_5$) | 6, 5614 |

We give each participant an authorization key (X, F(X)). Here, '$P_{x-1}$' instead of '$P_x$' is shared with the participant. Points start from (1, P(1)), not from (0, P(0)) because P(0) is the secret key value. The organization server maintains the Policy_DB based on the participant decryption points.

**3.4 Authorization Key Sharing with Asymmetric Encryption**

The secure sharing of a Secret key is performed based on the improved RSA technique.

*3.4.1 Encrypted Key Sharing*

The symmetric secret key ($PK_{SK}$) is shared for each participant using the public key technique. In this paper, the security of the secret key is achieved by encrypting it with Asymmetric Cryptosystem and shared with authorized users. The algorithm for the Improved RSA cryptosystem is as follows:

**Algorithm 3:** Key Generation using Improved RSA Cryptosystem

Step 1. The system generates two prime numbers, 'Pi' and 'Qi', for each $i^{th}$ user.
Step 2. Moduli 'ni' is calculated (ni = $\sum_{i=0}^{n} P_i \times Q_i$).
Step 3. Calculate $\Phi$ (n) = (P - 1) (Q - 1) and λ (n) = LCM (n) = LCM ($\Phi$ (n)), where 'λ' is the Totient function.
Step 4. Select (e), where the value 'e' lies in-between (e, 1 < e < $\Phi$ (n)), such that (e, $\Phi$ (n)) = 1).
Step 5. Compute [(Public Key ($Pub\_K_i$) × Private key ($Sec\_K_i$)) = 1 × mod × $\Phi(n)$] using CRT.
Step 6. Encrypt: $P_k = (R_n, PK_{SK})^{Pub\_K_i} \bmod n$.

Here '$PK_{SK}$' = Secret key, and '$R_n$' = randomized power variable of Involution Function-Based Stream Cipher. The Encrypted key ($P_k$) and Private Key ($Sec\_K_i$) is shared with the receiver and also stored on the organization server.

*3.4.2 Decryption Process*

This section performs the decryption of the encrypted key using the Improved RSA technique. The improved RSA uses the Chinese remainder technique to reduce the computation time of the decryption process. Let us consider $(n_1, n_2, n_3, …, n_k)$, are pairwise co-prime numbers and $(a_1, a_2, a_3, …, a_k)$ are positive integer values generated using Eq. (16).

$$a^{(\Phi(n))} = 1 \bmod n \qquad (16)$$

The standard solution of CRT is present in the form of (X= $X_0$ + kN), where 'k' and N = ($n_1 \cdot n_2 \cdot n_3 \cdot … \cdot n_r$) are integers; the solution is expressed as Eq. (17).

$$X \equiv \sum_{i=0}^{n} a_i \bar{X}_i N_i \equiv 1 \bmod N \qquad (17)$$

Here $\bar{X}_i$ is the modular inverse of '$X_i$' is calculated as Eq. (18).

$$\bar{X}_i Ni \equiv 1 \bmod N \qquad (18)$$

Then, $\left[N_i = \frac{N}{n_i}\right]$ is reduced as:

$$X = X_0 \bmod N \tag{19}$$

Here $[X_0 = a_i \bar{X}_i N_i]$, it completes the proof of CRT with RSA. The decryption process works on Euler's Totient function, denoted as $\Phi(n)$. It gives the number of integer's less than '$n$' and co-prime to '$n$'. If '$n$' is prime, the solution is represented as Eq. (20).

$$\Phi(n) = n\left(1 - \frac{1}{n}\right) \tag{20}$$

If '$n$' is prime, then $[C^{\Phi(n)} \equiv 1 \bmod n]$, and GCD $(c, n) = 1$ & $n$ = prime, therefore $[C^{(n-1)} \equiv 1 \bmod n]$, we use the CRT and Totient function for decryption is represented as Eq. (21).

$$\begin{cases} D_p = Sec\_K_i \bmod (p-1) \\ D_q = Sec\_K_i \bmod (q-1) \end{cases} \tag{21}$$

Here '$p$' and '$q$' are relatively prime's, reducing the computational time.

$$\begin{cases} X_p = (P_k)^{D_p} \bmod p \\ X_q = (P_k)^{D_q} \bmod q \end{cases} \tag{22}$$

Here, $P_k$ = Encrypted key that is used for Involution Function-Based Stream Cipher. The CRT-based RSA algorithm's unique solution is generated using Eq. (23).

$$M \equiv [(X_p \times q \times q^{-1}) + (X_q \times p \times p^{-1})] \bmod n \tag{23}$$

Here $p^{-1} = p \bmod q$ and $q^{-1} = q \bmod p$ are modulo inverse. It shows the theoretical steps involved in moduli reduction, and it is four times faster in computing than the basic RSA cryptosystem [25]. Now, every user requires an Encrypted Key $(P_k)$ to generate the $R_n$' = Randomized Power, and Private key $(Sec\_K_i)$ to access the encrypted file is expressed as Eq. (24).

$$\text{Decrypt } (PK_{SK}, R_n) = (P_k)^{Sec\_K_i} \bmod n \tag{24}$$

The Secret key has been recovered by decrypting the Encrypted key. The key generation uses the improved RSA technique to access the data from the root directory.

*3.4.3   Authorization Key Reconstruction*

The reconstruction of the key is performed on the organization server. The key reconstruction requires a minimum of three different authorization points (i) Organization Server Point, (ii) Owner point, and (iii) Receiver point to generate the scalar points over a parabolic curve. The advantage of the proposed model is that the organization server doesn't store any owner's points, and the contribution of all the parties in the proposed scheme controls the Denial-of-Service (DoS) attack and the decrypted points $(X, F(X))$ are used for interpolation. Let us consider $F(X_1, Y_1)$ = Organization server point, $F(X_3, Y_3)$ = Owner point, and $F(X_4, Y_4)$ = Receiver Point, and the respective values are (2, 1942), (4, 3402), and (5, 4414) for the key reconstruction. The Lagrange basic polynomial for point generation from a curve is defined as Eq. (25).

$$l_j(x) = \prod_{\substack{0 \leq m \leq k \\ m \neq j}} \frac{x - x_m}{x_j - x_m} \tag{25}$$

where $(0 \leq j \leq k)$, the assumption is that there are no two '$x_j$' are the same, where $(m \neq j)$ and $(x_j - x_m) \neq 0$. Now,

$$l_0(x) = \frac{x-x_1}{x_0-x_1} \times \frac{x-x_2}{x_0-x_2} \tag{26}$$

$$l_1(x) = \frac{x-x_0}{x_1-x_0} \times \frac{x-x_2}{x_1-x_2} \tag{27}$$

$$l_2(x) = \frac{x-x_0}{x_2-x_0} \times \frac{x-x_1}{x_2-x_1} \tag{28}$$

Now, calculate the $l_0(x)$, $l_1(x)$, and $l_2(x)$ to reconstruct the parabolic equation and retrieve the secret key. Therefore, construct the $l_0$, $l_1$, and $l_2$ is defined as Eq. (29).

$$f(x) = \sum_{j=0}^{2} y_i \times l_j(x) \tag{29}$$

The Lagrange interpolation required a minimum of three authorization points (Organization Server Point, Owner Point, and Receiver Point) to reconstruct the Secret key. The Secret key is used for the Functional-Based Stream Cipher to decrypt encrypted data.

## 4. PERFORMANCE AND SECURITY EVALUATION

In this paper, the proposed model uses the local organization server for encryption/decryption, key generation, and authorization key generation for every user. Data users are considered unreliable and acquire as much privilege as possible. Authorization is required to access and share files properly. The proposed model construction mainly focuses on scheme setup and key generation. The paper's subsequent section evaluates the performance of the proposed method and compares it to similar work.

### 4.1 Statistical Analysis

The proposed model uses the KM generator for random bit generation, later used for random padding and encryption processes. The random bit generation requires a Linux Mint platform, a 64-bit processor with a hardware specification of 4GB RAM, an Intel core i5, and a single-core processor using a CPU frequency of 1.70GHz. Our proposed approach involves utilizing an organizational server that enables users to upload their original data. The random bit is generated and stored in the matrix 'A'; the random sequence matches the original image's matrix B (256×256). The Functional-Based Stream Cipher uses bit-level encryption to encrypt the data. Finally, we evaluate the encrypted image pixel's uniformity using statistical analysis.

*4.1.1   Pseudorandom Number Generation*

This section present the random bit generation for cryptographic application.

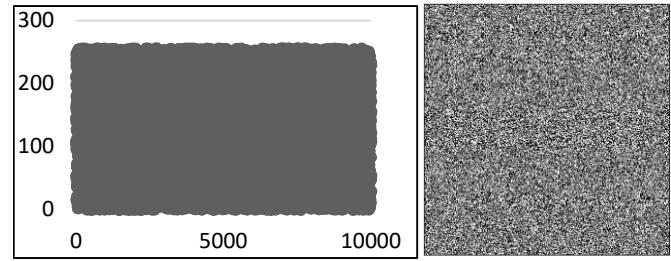

(a) Pictorial representation of random variable

(b) Pictorial pixel representation

**Figure 2. Pseudorandom Random Bit Generation**

Fig. 2 represents the random bit generation using KM-Generator, which is cryptographically secure and uniformly distributed throughout the region. Fig. 2(a) illustrates the binary output frame of (256×256) with randomness properties. Fig. 2(b) shows the histogram plot of the generated sequence in 256 window frames, where the pixel plot shows the uniform distribution of 0's and 1's and doesn't show any pattern in a generation. Here we examine the encryption of image files (.jpg) for encryption using Functional-Based Stream Cipher. The image file is in the format of (.jpg) and downloaded from the website (http://www.imageprocessingplace.com/root_files_V3/image_databases.htm, accessed on 23-4-2023).

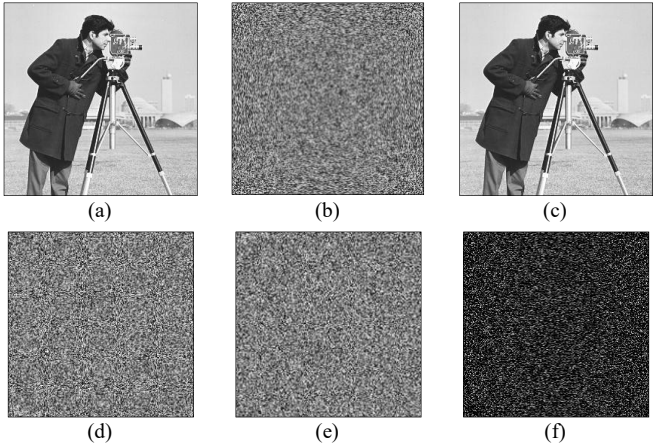

**Figure 3. Image Encryption using Functional-Based Stream Cipher**

Fig. 3 represents the encryption and decryption process of the cameraman's image with a different set of keys. Fig. 3(a) presents the original image. Fig. 3(b) illustrates the encrypted image using the symmetric secret key 'K1'. Fig. 3(c) represents the decrypted image using the symmetric secret key 'K1'. Now, the sensitivity of FBSC is evaluated by slightly changing the decryption key 'K2', presented in Fig. 3(d). The re-encryption process with 'K2' is presented in Fig. 3(e). The difference between the two different encryptions with 'K1' and 'K2' is present in Fig. 3(f). It observes that the proposed Function-Based Stream Cipher is extremely sensitive with the initial parameter for key generation and doesn't reveal any sensitive information. We execute the same operation with four different benchmark images (Lena, House, Boat, and Jet plane), each with a resolution of (256×256) pixels [29].

*4.1.2 NIST Statistical Suite*

This section presents the observation of pixel deviation in the encrypted image. The encrypted images are considered cryptographically secure and uniformly distributed once they pass all the NIST statistical tests. The input parameter for the test is (i) 128 bits of block length for the block frequency test, (ii) 9 bits for the non-overlapping and overlapping test, (iii) 500 bits for Linear Complexity test, (iv) 16 bits of block length for Sequential test, and (v) 10 bits length for approximation of entropy test [27].

| NIST Statistical Test Description |
|---|
| • The frequency test estimates the 0's and 1's ratio in the generated random stream. Suppose the proportion is not close to 50/50. In that case, it may indicate a deviation from randomness.
• Block Frequency test detects the amount of 1's in M-block sizes.
• The cumulative Sum test determines whether the sequence stays positive or negative for too long.
• Runs test determines the oscillation of 0's and 1's that are too fast or slow. It also examines the frequency of runs of consecutive 0's and 1's in the sequence.
• The longest run of 1's the block to examine the randomness in the generated sequence. It determines the length of the longest run of ones in each block of the sequence and compares it with expected values.
• The rank test checks the linear dependencies among fixed-length sub-strings of the original sequence. It also examines the ranks of specific matrices derived from the sequence.
• FFT test detects repetitive patterns close to each other. It applies the fast Fourier transform to the sequence and examines the resulting spectrum.
• The non-Overlapping Template Detection test detects the occurrences of non-periodic arrangements in generated sequence. It examines the frequency of a set of predefined templates in the sequence.
• The overlapping Template Matching test identifies the number of pre-specified target string occurrences. It examines the frequency of a set of overlapping patterns in the sequence.
• A Universal Statistical test detects the number of bits between matching patterns. It examines the distribution of the number of bits between pattern matches in the sequence.
• The approximate Entropy test compares the frequency of overlapping blocks of two consecutive lengths. It examines the similarity between the frequency distribution of the blocks of size 'm' and (m+1).
• Random Excursion test examines the number of cycles that return to zero for different cycle lengths.
• Random Excursion Variant test examines the cycles in the generated stream at every state.
• The linear Complexity test determines that the sequence is complex enough to be considered random. It measures the complexity of the sequence by counting the number of distinct linear feedback shift registers.
• Serial 1 test examines the frequency of all possible length patterns 'm' in the generated sequence.
• Serial 2 tests detect the random sequence's 2m bit overlapping patterns. It examines the frequency of all possible patterns of length 2m in the sequence. |

The NIST test applies to five benchmark encrypted images, and the histogram plot is used to represent the p-value of each test. The statistical hypothesis for the testing of encrypted images works on two possible outcomes $H_0$ = accept the uniformity of encrypted image pixels, and $H_1$ = reject the null hypothesis. Table 2 presents the hypothesis situation to accept the uniformity of pixels in an encrypted image.

**Table 2: Hypothesis Testing Scenario**

| Situation | Scenario 1 | Scenario 2 |
|---|---|---|
| Uniformity in encrypted image pixel ($H_0$) | $H_0$ is accepted No error | $H_0$ is rejected (Accept $H_1$) Type I error |

| Non-uniformity in encrypted image pixels ($H_1$) | Type II error | No error |

Type I and Type II errors consider the encrypted image's randomness. If the pixels are random, the null hypothesis is rejected, and concluded that it is non-random (Type I error). If the stream non-random accepts the null hypothesis, it determines that it is random (Type II error). Fig. 4 presents the p-value of the NIST test performed on encrypted data. The p-valve plot determines the randomness of encrypted image pixels. The fixed significance level ($\alpha > 0.01$) defines the randomness of pixels. The histogram plot of all 16 tests has a (p-value ($\alpha$) > 0.01), which indicates that the Functional-based encryption exhibits significant randomness and generates cryptographically secure data for storage and sharing.

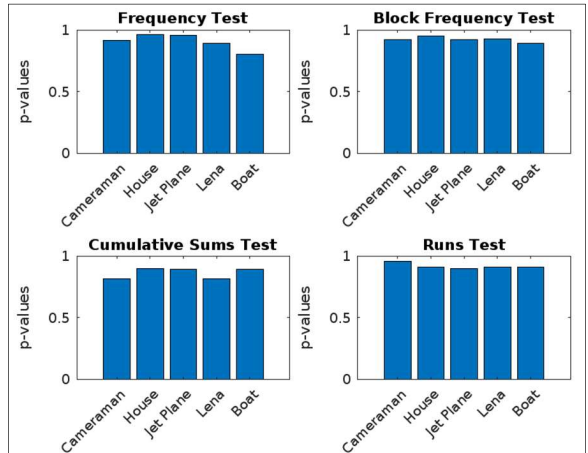

**a)** Histogram Plot of NIST Statistical Test Results: Frequency, Block Frequency, Cumulative Sum, and Runs Tests

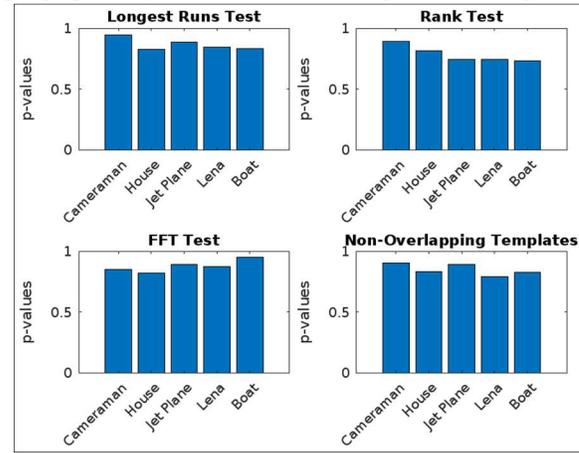

**(b)** Histogram Plot of NIST Statistical Test Results: Longest Runs, Rank, FFT, and Non-Overlapping Tests

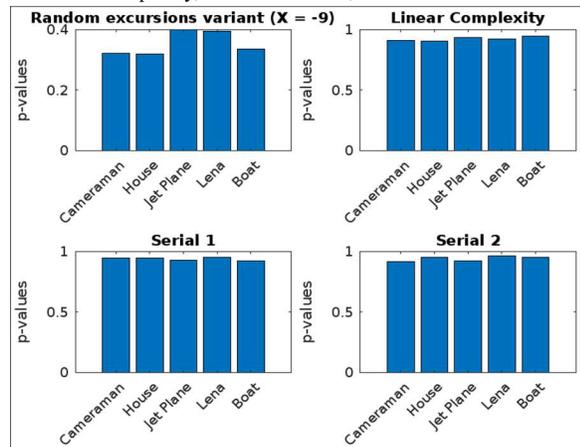

**(c)** Histogram Plot of NIST Statistical Test Results: Random Excursion Variant, Linear Complexity, Serial 1 and Serial 2 Tests

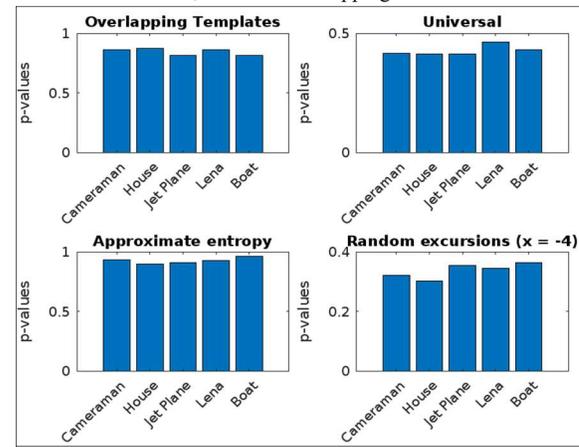

**(d)** Histogram Plot of NIST Statistical Test Results: Overlapping Templates, Universal, Approximate entropy, and Random excursions Tests

**Figure 4. NIST Statistical Test on Encrypted Images**

### 4.1.3 Histogram Analysis

This section compares the encrypted and original image pixels using a histogram plot. Table 3 presents the original image pixels distributed non-uniformly throughout the region, while the encrypted image shows a uniform distribution of pixels in the histogram graph plot.

**Table 3: Histogram Analysis**

| Original Image | Pictorial Original Image | Pictorial Encrypted Image |
|---|---|---|
| (boat image) | (histogram) | (histogram) |
| (cameraman image) | (histogram) | (histogram) |

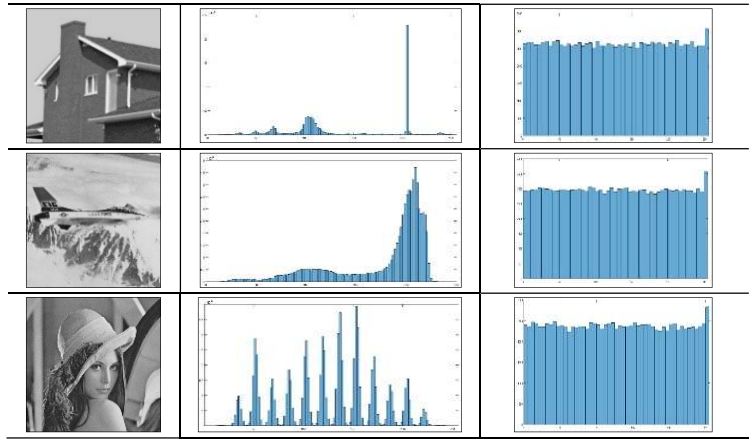

### 4.1.4 Correlation Coefficient (CF) Analysis

The correlation coefficient analysis uses the mean and variance of pixels within the nearest pixel and is computed as Eq. (30).

$$Corr\ (X,Y) = \frac{\left|\frac{1}{N}\sum_{i=1}^{n}(X_i-E(X))(Y_i-E(Y))\right|}{\sqrt{\frac{1}{N}\sum_{i=1}^{n}(X_i-E(X))^2 \times (Y_i-E(Y))^2}} \quad (30)$$

Here, E(x) and E(y) are the mean of the pixel value, and it is computed as Eq. (31).

$$E(X) = \frac{1}{N}\sum_{i=1}^{n} X_i\ \&\ E(Y) = \frac{1}{N}\sum_{i=1}^{n} Y_i \quad (31)$$

Table 4 summarizes the results of correlation coefficient analysis for five benchmark grayscale images of size (256×256). The test includes 16,384 pairs of neighboring pixels to perform the CF analysis. The correlation coefficient values for the original images are closer to 1, indicating a higher degree of correlation between neighboring pixels. In contrast, the correlation coefficient values for the encrypted data are more relative to 0, meaning a lower degree of correlation between neighboring pixels. The observation suggests that the proposed Involution Function-Based Stream Cipher generates high entropy among neighboring pixels.

**Table 4: Correlation Coefficient Analysis on various Images**

| Technique | Image Set (256 × 256) | Direction (Horizontal (H), Vertical (V), Diagonal (D)) Original Image | Direction (Horizontal (H), Vertical (V), Diagonal (D)) Encrypted Image |
|---|---|---|---|
| Proposed Function-Based Stream Cipher | Boat | H (0.9588) | H (-0.0025) |
| | | V (0.9625) | V (-0.0016) |
| | | D (0.9897) | D (-0.0005) |
| | House | H (0.9754) | H (-0.0017) |
| | | V (0.9832) | V (-0.0029) |
| | | D (0.9788) | D (-0.0025) |
| | Jet Plane | H (0.9764) | H (-0.0021) |
| | | V (0.9761) | V (-0.0014) |
| | | D (0.9765) | D (-0.0013) |
| | Lena | H (0.9519) | H (-0.0011) |
| | | V (0.9515) | V (-0.0012) |
| | | D (0.9893) | D (-0.0017) |
| | Cameraman | H (0.9862) | H (-0.0028) |
| | | V (0.9921) | V (-0.0025) |
| | | D (0.9615) | D (-0.0017) |
| DNA-BST [28] | Boat | H (0.9788) | H (0.0014) |
| | | V (0.9825) | V (0.0012) |
| | | D (0.9797) | D (0.0029) |
| | House | H (0.9854) | H (0.0033) |
| | | V (0.9732) | V (0.0021) |
| | | D (0.9760) | D (0.0022) |
| | Jet Plane | H (0.9764) | H (0.0035) |
| | | V (0.9861) | V (0.0036) |
| | | D (0.9998) | D (0.0022) |
| | Lena | H (0.9519) | H (0.0015) |
| | | V (0.9615) | V (0.0019) |
| | | D (0.9693) | D (0.0012) |
| | Cameraman | H (0.9662) | H (0.0014) |
| | | V (0.9821) | V (0.0014) |
| | | D (0.9715) | D (0.0022) |

### 4.2 Computation Overhead

This section presents the observation on the number of attributes used to construct the parabolic curve to provide the multiparty authorization key. Fig. 5 shows the encryption time of a file with multiple attributes. Encryption time increases when we add more attributes to the encrypted file. Fig. 6 presents the decryption cost on both the organization's server and user sides. The calculations include the cost of symmetric encryption and decoding, which is much more efficient than the asymmetric method and unchanged by errors. The cost of encryption and decryption increases linearly with an increase in the user attributes.

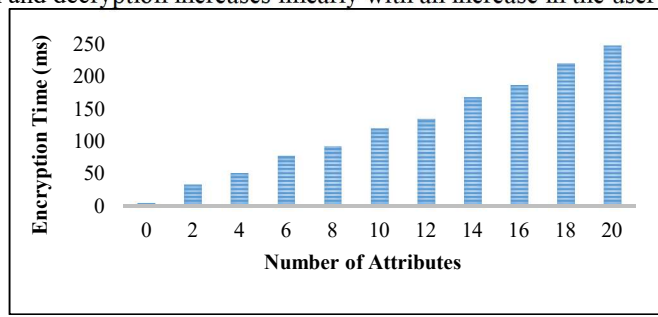

Figure 5. Encryption Time

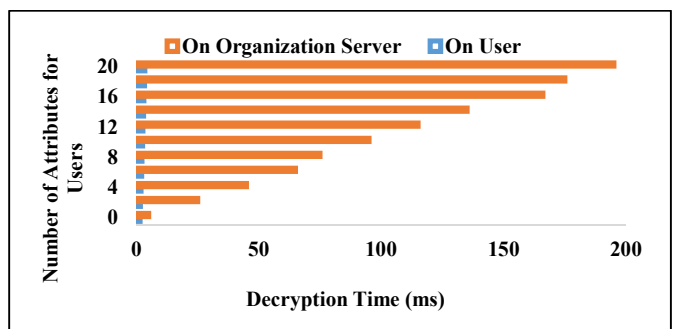

Figure 6. Decryption Time

### 4.3 Storage Overhead

This section represents the impact of attribute size on the storage overhead of an organization's server. The owner and user points are required for key reconstruction in the proposed system. The size of each parameter in the scheme is represented by $|P_i|$, while '$m$' and '$m_j$' denote the total number of attributes for owners and users, respectively. Table 5 compares the storage overhead of different parameter sizes used for key generation.

$m_{AA}$ = the number of elements in the management domain.
$K_c$ = the number of elements rooted in the ciphertext.

Table 5: Comparison of Storage Overhead

| Party | Proposed Model | Yang's DAC-MACS [17] |
|---|---|---|
| On User | $(m+1)\|P_i\|$ | $(m_{AA}+3)\|P_i\|$ |
| Organization Server | $(K_c+1)\|P_i\|$ | $(3K_c+3)\|P_i\|$ |

Our system stores the authorization point generated using a parabolic curve. The secret key is not stored on the organization server. The DAC-MACS model [17] utilizes the version number of each attribute and the secret authority key for key reconstruction. Our proposed scheme only requires the local organization server to store the authorization points as a token for key reconstruction. The secret key and encrypted file are deleted once encrypted data and authorization points are shared. The organization server manages all attributes, while the cloud server only stores the backup file of ACL. Therefore, the proposed system holds fewer parameters than Yang's DAC-MACS. If the owner's attributes exceed the user identities, then it will increase the storage overhead of data consumers (j) and affect the '$m_i$,' with the same '$m_j$'. The same user identities merge, reducing the storage overhead in this aspect. In other words, a user's storage overhead in our scheme is not affected by the users' identities.

### 4.4 Security Analysis

**Theorem 1:** In our system, if an attacker possesses '$j$' shares of the authorization key for one attribute, it is equivalent to the adversary having no secret key share.

**Proof:** To reconstruct the key in our proposed system, we require a minimum of three distinct points: the Organization Server Point, Owner Point, and Receiver Point, to generate the parabolic curve. Let us assume that '$P_i$' denote the probability of adversary guessing an available secret attribute key with '$i$' shares. The authentication is performed based on the Owner point, meaning that an adversary cannot access the encrypted file unless the data owners don't provide the authorization points to the system [30]. It involves two requirements (i) the set of attribute points must satisfy encryption, and (ii) the data owner must approve the access of the data file containing the attributes and share the key.

**Theorem 2:** Our system has been designed with a robust security feature to resist collusion attacks.

**Proof:** In our proposed system, the attributes for each data consumer are User_Id ($U_i$) and User_Type ($UT_i$), and User_Credentials ($UC_i$). The $UC_i$ is embedded in a parabolic equation with secret keys generated from functional-based encryption. The Organization Server Point and Owner Point are necessary to access the encrypted file, allowing only one recipient to access the file at a time. This mechanism safeguards against unauthorized access by preventing the attachment of the owner point to generate secret keys. As a result, if two or more consumers cannot complete the task independently, they cannot jointly decrypt the ciphertext.

**Theorem 3:** Our system maintains security even when some owner points are compromised.

**Proof:** In related schemes such as [17, 23], the scheme constraints that Organization Server Point, Receiver point, and Owner point can only generate secret keys. This mechanism works under the security assumption if any Receiver point is compromised, and the adversary uses that point for the key generation. Here, the proposed system is a two-way authentication where the organization server authenticates the adversary credentials. Once completed, the organization server requires the Owner Point to generate the secret key. Now the data owner validates the receiver using the access control list. With the above properties, the construction of our scheme can still work even if some owner points are compromised.

## 5. CONCLUSION AND FUTURE WORK

This paper uses ABE based ACL Scheme and Functional-Based Stream Cipher techniques for authorization and secure data storage. Function-based encryption provides a secure and efficient way to encrypt and store data on the organization's server. The traditional schemes require huge computation time and are prone to data leakage. The proposed model enables the authorization process and access control scheme to access the encrypted data. Our scheme includes a specified threshold value (T) that legally authorized users must meet before distributing attribute-associated keys that allow for the decryption of confidential data. The Lagrange interpolation technique generates secret tokens and shares them with privileged users. The access privilege tokens are used to reconstruct the user's decryption key. Privileged users decrypt the encrypted data in the proposed model without sharing confidential information. The randomness and deviation in encrypted image pixels are detected using statistical analysis (NIST and Correlation Coefficient). The security and performance analysis show that the proposed scheme is highly secure and robust against various threats. In the future, we plan to implement this model in various real-life applications, including social networking, medical, and military systems, to create a safe environment for storing and transmitting multimedia data.